\documentclass[3p,times,procedia]{elsarticle}
\usepackage{nupha_ecrc}
\usepackage{url}
\usepackage{hyperref}

\volume{00}

\firstpage{1}

\journalname{Nuclear Physics A}

\runauth{}

\jid{nupha}

\jnltitlelogo{Nuclear Physics A}

\usepackage{amssymb}

\usepackage[figuresright]{rotating}
\usepackage{float}

\newcommand{\trento}{T$\mathrel{\protect\raisebox{-2.1pt}{R}}$ENTo}

\begin{document}

\begin{frontmatter}



\dochead{XXVIIth International Conference on Ultrarelativistic Nucleus-Nucleus Collisions\\ (Quark Matter 2018)}

\title{Geometric scaling in symmetric nucleus-nucleus collisions}


\author[label1,label2]{Rudolph Rogly}
\author[label1]{Giuliano Giacalone}
\author[label1]{Jean-Yves Ollitrault}

\address[label1]{Institut de physique th\'eorique, Universit\'e Paris Saclay, CNRS,
CEA, 91191 Gif-sur-Yvette, France}
\address[label2]{MINES ParisTech, PSL Research University, 60 Boulevard Saint-Michel, 75006 Paris, France}

\begin{abstract}
  We show that the centrality dependence of the multiplicity is identical in Pb+Pb and Xe+Xe collisions at the LHC, up to a constant factor.
  This geometric scaling is revealed if one defines centrality according to impact parameter, as opposed to the usual experimental definition, which is in terms of multiplicity. 
  We reconstruct the impact parameter dependence of the multiplicity from experimental data using a recently-developed inversion method, 
which turns out to describe ALICE Xe+Xe multiplicity data much better than usual Monte Carlo Glauber fits.
The multiplicity as function of impact parameter extracted from ALICE data is compared to model calculations. 
\end{abstract}


\end{frontmatter}


\section{Introduction}
\label{s:intro}

The impact parameter, $b$, of a nucleus-nucleus collision at ultrarelativistic energy is a well-defined quantity, in the sense that its quantum uncertainty is negligible. 
It is a key input to model calculations, such as hydrodynamic calculations.
However, $b$ cannot be directly measured.
Therefore, measuring a quantity as simple as the average multiplicity at fixed $b$ is not straightforward. 
This is unfortunate, as such a quantity is actually very useful: 
It is typically used to fix the initial temperature~\cite{Luzum:2008cw} or entropy~\cite{Noronha-Hostler:2013gga} in hydrodynamic calculations. 

We show that this quantity can be extracted from data under general assumptions. 
The impact parameter is related to the multiplicity, in the sense that more central collisions produce on average more particles, but the relation between multiplicity and impact parameter is probabilistic, not one to one. 
We have shown that this relation can be inferred from data~\cite{Das:2017ned,Rogly:2018ddx} without relying on any specific microscopic model. 
The problem of reconstructing impact parameter from multiplicity is a typical inverse problem, which is solved by a deconvolution method. 
The main ingredient is the fluctuation kernel which is used to model multiplicity fluctuations at a fixed impact parameter.
This kernel is presented in Sec.~\ref{s:gamma}.
We then recall briefly the principle of the inversion method and apply 
it to recent Xe+Xe data at $5.44$~TeV in Sec.~\ref{s:xexe}.
In Sec.~\ref{s:scaling}, the outcome of our procedure is used to perform a comparison of the variation of the mean multiplicity as a function of the true centrality (defined by the impact parameter) between Xe+Xe and Pb+Pb collisions.
The resulting variation of multiplicity is then compared to model calculations. 

\section{Parametrizing multiplicity fluctuations: negative binomial and gamma distributions}
\label{s:gamma}

\begin{figure}[t!]
\begin{center}
\includegraphics[width=0.68\linewidth]{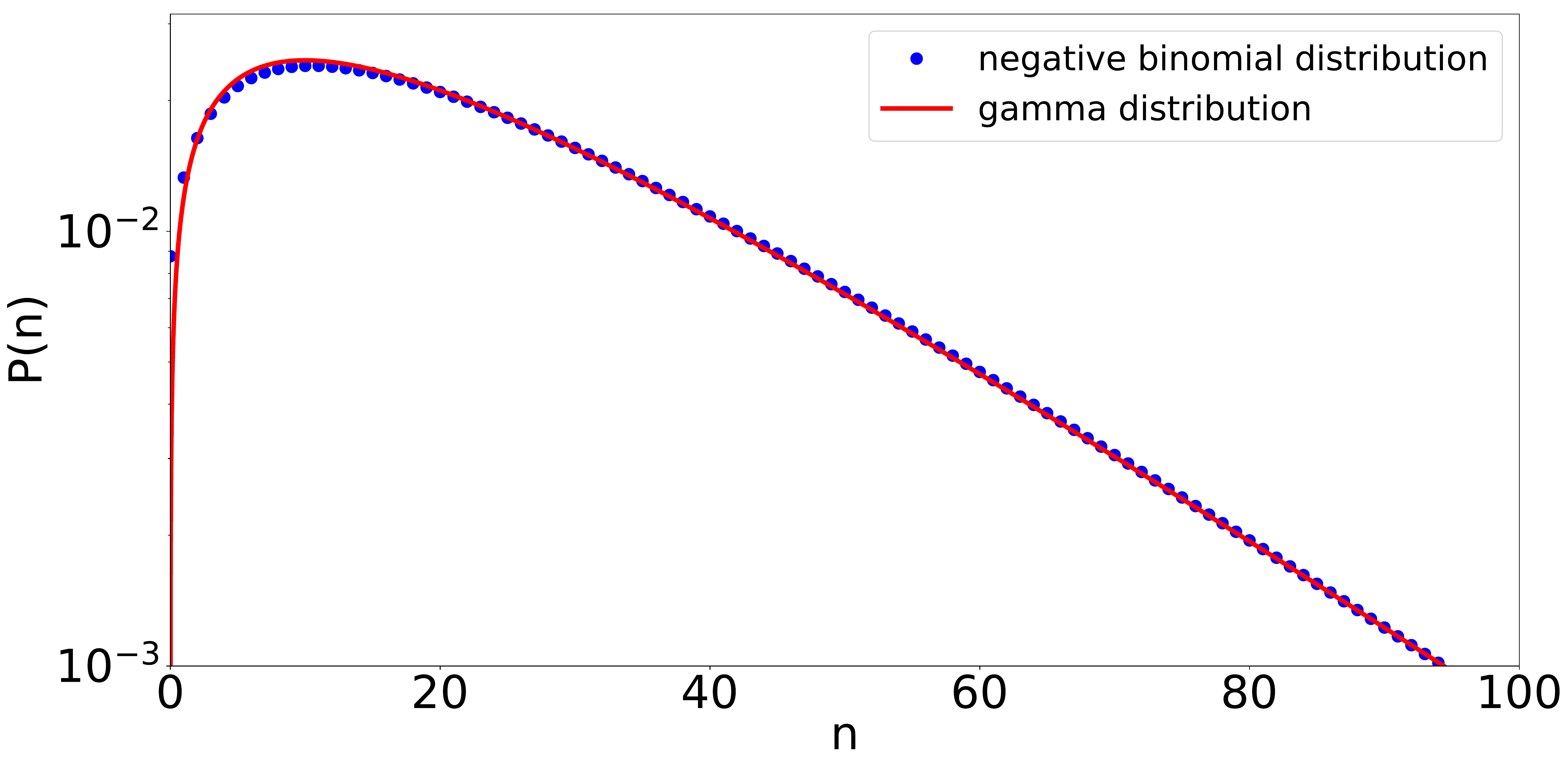} 
\end{center}
\caption{(Color online) 
\label{fig:kernel}
Symbols: Negative binomial distribution used by the ALICE collaboration in their Glauber Monte Carlo fit to Pb+Pb data~\cite{Abelev:2013qoq}. Solid line: gamma distribution with the same mean and variance.
Note that Glauber fits typically assume NBD fluctuations of the multiplicity for a fixed number of ancestors, while we assume gamma fluctuations for a fixed impact parameter.
}
\end{figure}
The multiplicity of particles, $n$, seen in a detector fluctuates event to event for fixed $b$. These fluctuations are partly due to dynamical fluctuations, and to statistical fluctuations which depend on the detector acceptance.
The principle of our method is to parametrize these fluctuations---or, more precisely, the probability of $n$ at fixed $b$, denoted by $P(n|b)$---in the most general way.
If fluctuations are small, one expects them to be Gaussian~\cite{Das:2017ned}. Since the multiplicity is positive, however, a more accurate parametrization of its fluctuations is provided by the gamma distribution~\cite{Rogly:2018ddx} which, unlike the Gaussian, has positive support. 
The gamma distribution~\cite{Broniowski:2007nz,Moreland:2014oya} can be viewed as a continuous version of the negative binomial distribution (NBD), which has long been used to fit multiplicity distributions in high-energy collisions~\cite{Giovannini:1985mz}.
We choose the continuous version because it also applies to the case when one uses an energy, instead of a multiplicity, to estimate the centrality. 
The similarity between the gamma distribution and the NBD is illustrated in Fig.~\ref{fig:kernel}.

\section{Inversion method and application to Xe+Xe collisions at $\sqrt{s_{NN}}=5.44$~TeV} 
\label{s:xexe}

The observed multiplicity distribution, $P(n)$, is obtained by summing the contributions to multiplicity at all impact parameters:
\begin{equation}
  \label{convolution}
P(n)=\int_0^{\infty} P(n|b)P(b)db=\int_0^1 P(n|c)dc, 
\end{equation}
where $P(b)$ is the probability distribution of impact parameter, and $c$ denotes the centrality, defined as the cumulative distribution of impact parameter: $c\equiv\int_0^bP(b')db'$. 
We write $P(n|c)=P(n|b)$ the probability of $n$ at fixed centrality, that is, at fixed $b$. 
As explained above, $P(n|c)$ is assumed to be a gamma distribution.
The gamma distribution has two parameters corresponding to the mean, $\bar n$, and to the variance, $\sigma^2$. Both depend on $c$. 
We assume that $\bar n$ is a smooth, monotonously decreasing function of $c$.
We also assume for simplicity that the ratio of the variance to the mean, $\sigma^2/\bar n$, is independent of $c$, but the results presented below are robust with respect to this assumption~\cite{Broniowski:2001ei}.
We then fit $\bar n(c)$ and $\sigma^2(c=0)$ so that $P(n)$ defined by Eq.~(\ref{convolution}) matches experimental data.
In practice, we shall take $\bar n$ as the exponential of a polynomial.
The exponential guarantees positivity, and the degree of the polynomial (i.e., the number of fit parameters) is chosen large enough to yield a perfect fit of data (below, a polynomial of degree 6 is used).
\begin{figure}[t!]
\begin{center}
\includegraphics[width=0.87\linewidth]{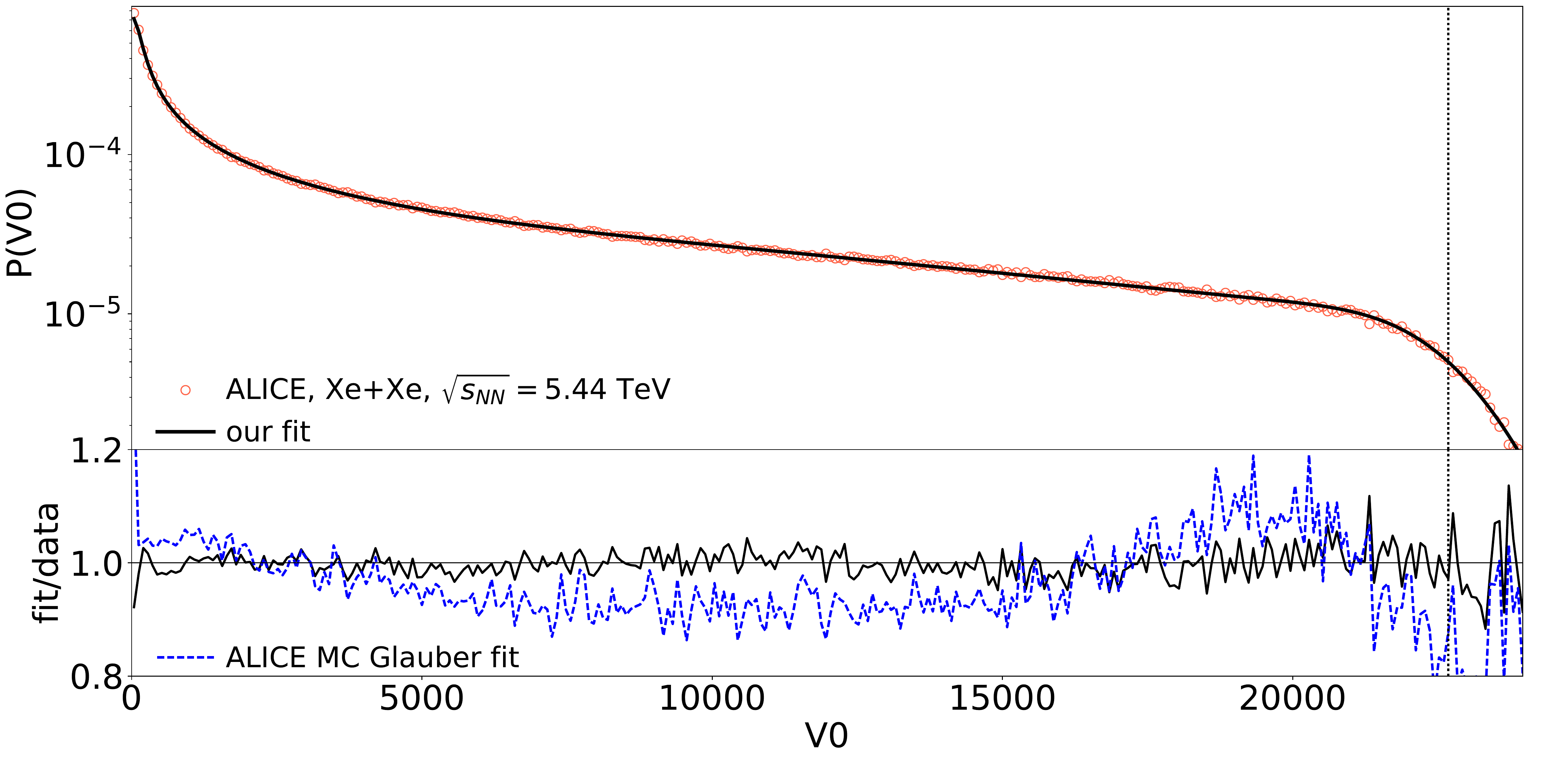} 
\end{center}
\caption{(Color online) 
\label{fig:ALICE}
Top panel: circles: ALICE data for the distribution of the multiplicity in the V0 detector~\cite{beomkyu_kim}. Solid line: fit using Eq.~(\ref{convolution}) (see text), where $n$ denotes the V0 amplitude. The vertical line is the position of the knee, defined as the mean value of V0 for $b=0$. Bottom panel: ratio fit over data for our fit (solid line) and for the MC Glauber fit used by ALICE (dashed line). 
}
\end{figure}

We now apply this procedure to ALICE data on Xe+Xe collisions at $\sqrt{s_{\rm NN}}=5.44$~TeV~\cite{beomkyu_kim}. 
In the case of ALICE data, $n$ is the multiplicity of hits in the V0 detector, which we shall simply denote by V0. 
We exclude values of $V0$ below the ``anchor point''~\cite{Abelev:2013qoq} where a fraction of events are missed.
We normalize $P(V0)$ in such a way that the fraction of events above the anchor point matches the value specified by ALICE.
We fit the measured $P(V0)$ using Eq.~(\ref{convolution}). 
Our fit is displayed together with ALICE data in the top panel of Fig.~\ref{fig:ALICE}. 
The bottom panel displays the ratio fit/data.
This ratio is very close to 1 for our fit, but deviates by up to 10\% for the fit performed by the ALICE collaboration using a Monte Carlo Glauber model~\cite{Loizides:2014vua}. 
This shows that our inversion method provides a more accurate representation of data than specific models of the collision. 
Note that our procedure neglects the deformation of the $^{129}$Xe nucleus.
Deformation is important for elliptic flow~\cite{Giacalone:2017dud,Eskola:2017bup}, but plays a modest role for multiplicity distributions~\cite{Moreland:2014oya}. 
This is confirmed by the excellent quality of our fit. 

\section{Comparison between Xe+Xe and Pb+Pb collisions}
\label{s:scaling}

\begin{figure}[t!]
\begin{center}
\includegraphics[width=\linewidth]{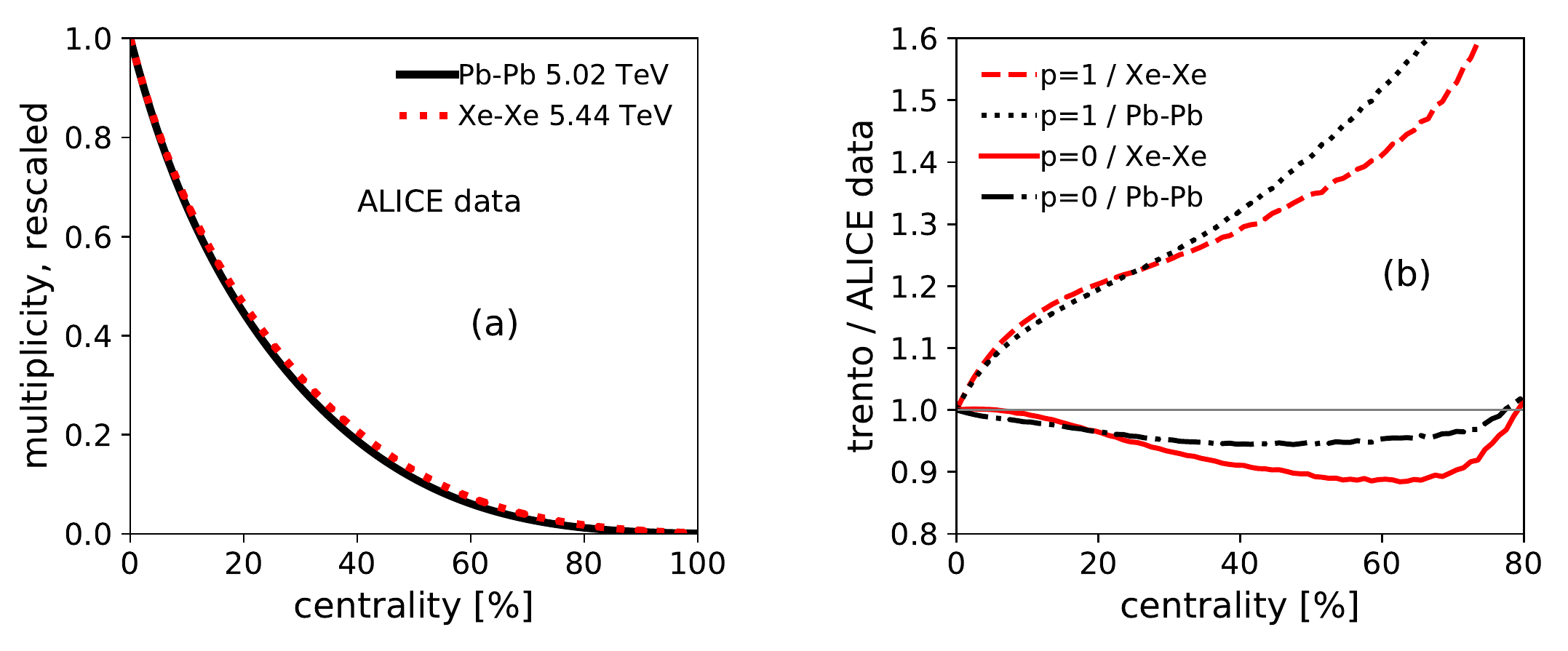} 
\end{center}
\caption{(Color online) 
\label{fig:scaling}
(a) Mean multiplicity versus centrality, rescaled by the value at $b=0$, for Pb+Pb collisions  at $\sqrt{s_{NN}}=5.02$~TeV (dashed line) and Xe+Xe collisions at $\sqrt{s_{NN}}=5.44$~TeV (solid line).
(b) Ratios model/data for \trento{}  with $p=1$ and $p=0$.
}
\end{figure}

Figure~\ref{fig:scaling}~(a) displays the mean multiplicity as a function of centrality percentile, as given by our fit to data. 
We again emphasize that this is the true centrality, defined from impact parameter (called ``$b$-centrality'', and denoted by $c_b$, in Refs.~\cite{Das:2017ned,Rogly:2018ddx}), which differs from the usual centrality, as defined experimentally from the cumulative distribution of the multiplicity. 
On the vertical axis, we divide the mean multiplicity by its value at $b=0$. 
When plotted in this way, Xe+Xe and Pb+Pb collisions essentially fall on the same curve.
This means that colliding larger nuclei amounts to multiplying the multiplicity by a constant factor, which is independent of centrality.
We call this phenomenon geometric scaling for obvious reasons.
(It should not be confused with the geometric scaling observed in deep inelastic electron-proton scattering~\cite{Stasto:2000er}.) 
It would be interesting to measure the proportionality factor between two different colliding systems at the same energy, and check whether the multiplicity is proportional to the mass number of colliding nuclei. 

We now show how the quantity displayed in Fig.~\ref{fig:scaling}~(a) can be used to test models.
One can calculate the same quantity in a model of the collision.
We test the \trento{} model~\cite{Moreland:2014oya}, which generates initial conditions for hydrodynamic calculations, and assume that the final multiplicity of an event is the total entropy calculated in \trento{}. 
The \trento{} model has a parameter $p$ which determines how the multiplicity depends on the thickness functions $T_A$ and $T_B$ of colliding nuclei.
We test the choices $p=1$ and $p=0$, which correspond respectively to $T_A+T_B$ 
(equivalent to the wounded nucleon model~\cite{Bialas:1976ed}, that is, to the Glauber model with participant scaling~\cite{Miller:2007ri}) 
and $\sqrt{T_AT_B}$ (similar to models inspired by high-energy QCD~\cite{Schenke:2012wb,Moreland:2014oya}). 
Fig.~\ref{fig:scaling}~(b) displays the ratio of model over data, where data are from panel (a). 
With both values of $p$, geometric scaling is satisfied to a good approximation, in the sense that Xe+Xe and Pb+Pb results are close. 
This is not surprising as scale invariance is one of the starting assumptions in the \trento{} model.
Note that scale invariance, on the other hand, is not satisfied by the two-component Glauber model (it is broken by the term proportional to the number of collisions).
The agreement of the \trento{} model with data is better with $p=0$ than with $p=1$, as already noted in the context of flow studies~\cite{Giacalone:2017uqx}.

The \trento{} model also implements gamma fluctuations of the multiplicity.
These fluctuations are switched off in the calculations presented in Fig.~\ref{fig:scaling} (b), but we have checked explicitly that results are essentially unchanged if fluctuations are included. 
This is not surprising, since we evaluate the mean multiplicity at a fixed impact parameter, which is averaged over fluctuations. 

In conclusion, we have applied a Bayesian inversion method to infer from data the variation of the mean multiplicity as a function of the true centrality, defined as a function of impact parameter.
We have shown that this variation is essentially independent of the nuclear size for symmetric collisions between large nuclei, thus revealing a geometrical scaling.
This simple observable allows to constrain models of particle production.

\section*{Acknowledgements}
We thank Alberica Toia for useful discussions, and for sending us ALICE data.

\bibliographystyle{elsarticle-num}
\bibliography{qmbib}


\end{document}